# Energy dissipation in graphene field-effect transistors

Marcus Freitag*, Mathias Steiner, Yves Martin, Vasili Perebeinos,

Zhihong Chen, James C. Tsang, Phaedon Avouris

IBM TJ Watson Research Center, Yorktown Heights, NY 10591

* email: mfreitag@us.ibm.com

## Abstract

We measure the temperature distribution in a biased single-layer graphene transistor using Raman scattering microscopy of the 2D-phonon band. Peak operating temperatures of 1050 K are reached in the middle of the graphene sheet at 210 KW cm$^{-2}$ of dissipated electric power. The metallic contacts act as heat sinks, but not in a dominant fashion. To explain the observed temperature profile and heating rate, we have to include heat-flow from the graphene to the gate oxide underneath, especially at elevated temperatures, where the graphene thermal conductivity is lowered due to umklapp scattering. Velocity saturation due to phonons with about 50 meV energy is inferred from the measured charge density via shifts in the Raman G-phonon band, suggesting that remote scattering (through field coupling) by substrate polar surface phonons increases the energy transfer to the substrate and at the same time limits the high-bias electronic conduction of graphene.

Graphene, the recently isolated 2-dimensional carbon material with unique properties due to its linear electronic dispersion, [1,2] is being actively explored for electronic applications. [3-8] Figures of merit that have attracted attention are the high mobilities reported especially in suspended graphene, [9, 10] the fact that graphene is the ultimately thin material, the stability of the carbon-carbon bond in graphene, the ability to induce a bandgap by electron confinement in graphene nanoribbons, [5-7] and its planar nature, which generally allows established patterning and etching techniques to be applied. Among the disadvantages are difficulties in fabricating wafer-sized graphene, the poor on/off current ratios, and the apparent degradation of carrier mobility in graphene once it is placed in contact with an oxide. Considering that graphene has been studied for just a few years, it is understandable that many important questions remain unanswered. One such question concerns heat generation and dissipation in graphene field-effect transistors (FETs). This would become an important issue for example when a large current drive is needed to address several other FETs, or when high switching speeds for radio-frequency applications are desired. [8] The dissipated electric power can then raise the operating temperature to a point where thermal management becomes critical.

To our knowledge, there is no published work on graphene thermal transport in field-effect transistors. Thermal transport properties of the related carbon nanotube field-effect transistors have previously been analyzed, using the break-down method, where the drain voltage is ramped up until devices fail. [11, 12] The oxidation temperature where breakdown occurs is roughly known, so a temperature could be assigned to the location where the device failed. In a different approach, the temperature distribution in a biased multi-walled carbon nanotube was measured with a scanned thermal probe. [13] Raman spectroscopy has also been used to measure the population of specific phonons in nanotubes albeit without spatial information. [14-15] Here we use spatially-resolved Raman spectroscopy to measure the temperature distribution in biased graphene FETs. The position of the Raman-active 2D-phonon band near 2700cm$^{-1}$, involving two optical phonons with equal and opposite momentum near the zone boundary, is strongly dependent on the local temperature, and can be used as a microscopic thermometer. [17] The Raman G-optical phonon band at ~1600cm$^{-1}$ is, in addition to temperature, highly susceptible to local doping, [18, 19] and can be used to detect drain-bias induced changes in the Fermi level. We find that when the graphene FET is electrically driven to the current saturation regime, the 2D-derived temperature can reach 1000K. Direct heat flow from the graphene to the substrate dominates even in devices with thick (~300nm) gate oxides. We find that the interface thermal resistance to the underlying SiO$_2$ ($r_{Gr-SiO2} \sim 4 \cdot 10^{-8}$ K m$^2$/W) is smaller than might be expected for an oxide film coupled weakly to graphene. This we can explain in terms of direct energy transfer from hot conduction electrons to substrate polar surface phonons (field coupling).

The device discussed here is fabricated by mechanical exfoliation of graphene. [4] Contacts are formed on top of the graphene sheet by e-beam lithography, metal evaporation and liftoff. Metallic contacts consist of 0.5 nm Ti, 30 nm Pd, and 20 nm Au. The graphene FET is 2.65 μm long and W=1.45 μm wide. The SiO$_2$ gate oxide thickness is 300 nm. The FET can support electrical power densities up to at least 210 KW cm$^{-2}$. It shows p-type behavior with only hole conduction within the applied gate voltage range of ±10V (see supplemental materials). We performed in situ measurements of the Raman 2D and G-phonon band under bias. A 100x objective with numerical aperture of NA=0.8 was used to focus the $\lambda = 514.5$ nm laser beam onto the sample. Laser power levels on the sample were kept around 100 μW, where little laser-induced heating (<50K) can be observed. Raman-scattered light was collected with the same objective and analyzed with a Triax 322 (Horiba Jobin Yvon) spectrometer and a liquid-nitrogen cooled CCD. During the high-bias measurements, clean, dry nitrogen was flown over the FET to prevent oxidation. The position and intensity of both 2D and G-phonon bands are gate-voltage independent within our resolution of 1 cm$^{-1}$. Electrostatic gating effects, such as the ones reported in references [18,19], therefore play a minor role in the results below, probably due to the thick silicon-oxide used.

Figure 1 shows the 2D-band of a graphene FET while a current is flowing through it. The 2D energy decreases sharply with increasing electric power and the peak broadens. As Fig. 1b shows, the decrease in energy is roughly proportional to the dissipated electric power, which suggests that Joule heating is responsible for the phonon softening. Temperature-dependent measurements of the graphene 2D-phonon band have shown that its energy decreases linearly with temperature. [17] Using the proportionality factor of $-29.4$ K/cm$^{-1}$ from this reference, we can calibrate a temperature scale for our electronic measurements and in this way find that the center of the graphene FET is heating up at a rate of 3.3 K/(kW cm$^{-2}$). At the highest power density, 210 KW cm$^{-2}$, the graphene 2D-derived temperature reaches 1050K. From our experience with a number of other graphene FETs this is close to the maximal electrical power density that can be applied before devices start to fail, so the temperature calibration seems reasonable. Strain can also produce shifts in Raman bands [20-22] and one has to be aware of strain occurring in heating experiments. The amount of strain should however be similar in our electrically biased devices and in the thermally heated devices of reference [17]. The relation between Raman shift and temperature [17] should therefore hold in our experiments. The softening of the 2D-phonon band is especially useful for temperature measurements in graphene, since the G or 2D anti-Stokes intensities are very low. Indeed we tried to measure the anti-Stokes G-phonon band at the highest power density of 210 KW cm$^{-2}$ and were unsuccessful. It is interesting to note that in the case of carbon nanotubes, even though they contain two orders of magnitude fewer carbon atoms within a focused laser spot than graphene, the anti-Stokes G-phonon band can be enhanced sufficiently by exploiting exciton resonances. [14, 15]

In Fig. 1c, color-coded images of the 2D-band energy are plotted as a function of the lateral position of the laser spot on the FET. The energies can be converted to temperatures, as shown in the scale bar. Four of these images are displayed at 4 different drain voltages. The 2D-band temperature is hottest close to the center of the FET. At the edges of the graphene sheet, the 2D-band temperature is 15% reduced (Fig. 2b), while near the contacts it is close to 50% below that at the peak (Fig. 2a). The decrease in temperature near the edge of the graphene sheet is due to the lateral heat transport in the gate oxide and possibly also due to a reduction in current flowing near the edges of the FET due to edge doping. [23] The small deviations from perfect mirror symmetry can be attributed to differences in contact widths left (1.45 μm) and right (1.25 μm). As a result, the temperature maximum within the graphene sheet is not reached exactly in the middle of the FET, but instead 300 nm, or about 10% of the device length, to the right. The thermal conductivity of a single graphene layer has recently been measured using laser-heating of suspended graphene. [24, 25] The reported values, $4840-5300$ Wm$^{-1}$K$^{-1}$ and $3080-5150$ Wm$^{-1}$K$^{-1}$ at room temperature, are many orders of magnitude greater than the thermal conductivity of the underlying SiO$_2$ (~1 Wm$^{-1}$K$^{-1}$) and of course that of air ($0.025$ Wm$^{-1}$K$^{-1}$). Lateral heat flow within the one-atom thin graphene sheet should therefore be very important. The observed peaked temperature profile in the biased graphene FET is consistent with this simple picture.

For a quantitative analysis of the heat dissipation in the FET we used the thermal simulation software, PDEase from Macsyma Inc., which solves the heat diffusion equation

$$\text{div}(\kappa(x,z) \cdot \nabla T(x,z)) = 0$$

in 2 dimensions for a given geometry (see Figure 2). Here $\kappa(x,z)$ is the thermal conductivity of the various materials involved, $T(x,z)$ is the local temperature, $x$ is the direction along the FET, and $z$ is the direction into the oxide. The electric power is modeled as a uniform heat-flux (in W/area) deposited at the surface of the graphene layer. The device dimensions are all known, as are the thermal conductivities of the gate stack and the metallic contacts. The metal-graphene interface is modeled as an intimate contact (thermal resistance set to zero), while the important thermal interface resistance between graphene and the underlying SiO$_2$, $r_{Gr-SiO2}$, is an adjustable parameter. To compare the model with the experiment we convolute the calculated temperatures with a laser spot of 500nm width. This is necessary because the FET is only a few spot diameters long. The laser spot size of 500nm was determined from the falloff in Raman 2D-band intensity as the laser spot is moved across the graphene sheet. The graphene below the 0.5 μm wide and ~50nm thick metal contacts is not visible in the Raman measurement, so the calculated temperature there is not used for the convolution.

The thermal conductivity of graphene is experimentally known to be on the order of $\kappa_{Gr} = 5000$ Wm$^{-1}$K$^{-1}$ at room temperature. [24, 25] However, the temperatures reached in our experiments are significantly elevated, and the temperature dependence of the thermal conductivity has to be accounted for. Electronic contributions to the thermal conductivity of graphene are negligible compared to phonon contributions. [24] At low temperature the major phonon scattering mechanism is defect scattering, which is temperature independent, so the temperature dependence of the thermal conductivity is given by the number of phonon modes that are populated. The Debye temperature, up to which it is possible to populate higher phonon modes, is very high (thousands of degrees) in all sp$^2$ carbon materials due to the high energy of the optical phonons associated with the carbon-carbon bond, so the thermal conductivity can be expected to reach very high values in the absence of other scattering mechanisms (the ballistic limit). [26] At sufficiently high temperatures however, anharmonic (phonon-phonon) scattering sets in, and the thermal conductivity is dominated by the reduction in the mean free path due to umklapp scattering. [27, 28] In an umklapp scattering event, two incoming phonons with sufficiently large wave-vector create an outgoing phonon with wave-vector outside the first Brillouin zone. The outgoing phonon is related by a reciprocal lattice vector to a phonon inside the first Brillouin zone whose momentum is less than the total momentum of the incoming phonons. The phonon contribution to the thermal conductivity is therefore reduced.

Molecular dynamics simulations have shown that the high-temperature thermal conductivity of graphene is indeed dominated by umklapp scattering and falls off rapidly. [27] We are not aware of any published experimental work on the thermal conductivity of graphene at elevated temperatures. Fortunately, the graphene thermal conductivity can be seen as the large-diameter limit of the thermal conductivity of carbon nanotubes, [28] and for those, the onset of umklapp scattering is known to occur at T=350K. [29] We therefore make the assumption that the graphene thermal conductivity also drops above 350K and can be approximated as

$$\kappa_{Gr}(T) = \frac{5000 \text{ Wm}^{-1}\text{K}^{-1}}{1+0.01(T-350\text{K})}$$

(see supplemental information), which fits the slope beween 350K and 380K in reference [29], when scaled to the peak value of $\kappa_{Gr} = 5000$ Wm$^{-1}$K$^{-1}$. Below 350K we use the established value of $\kappa_{Gr} = 5000$ Wm$^{-1}$K$^{-1}$. [24,25] Within this assumption, the graphene thermal conductivity is reduced from $\kappa_{Gr} = 5000$ Wm$^{-1}$K$^{-1}$ to 850 Wm$^{-1}$K$^{-1}$ for temperatures between $T = 300$ K and 800 K. Our results below are not very sensitive to the exact functional form of the decrease in thermal conductivity at high temperatures, but they are clearly inconsistent with a constant thermal conductivity of $\kappa_{Gr} = 5000$ Wm$^{-1}$K$^{-1}$ or with an increasing thermal conductivity as would be the case in the ballistic limit.

In Figures 2c and 2d we show the modeled temperature distribution for three different electrical power levels and the comparison to the experimental 2D-derived temperatures. The best results are obtained for a thermal interface resistance of $r_{Gr-SiO2} = 4.2 \cdot 10^{-8}$ K m$^2$/W, which is equivalent to an additional SiO$_2$ layer of 42 nm thickness. Lateral heat flow in the thin graphene sheet is 5x larger then the lateral flow in SiO$_2$ (despite being 1000x thinner than the SiO$_2$ film). This helps to better spread the hot spot in the graphene sheet which would otherwise be even hotter in the middle, and it also spreads some of the heat power to the contacts. Nevertheless, 77% of the power is dissipated through the SiO$_2$ directly below the FET, while the remaining 23% are dissipated through the contacts and the neighboring graphene (see supplemental materials). Eventually, all the heat dissipates through the silicon substrate. According to our model, a measurable increase in temperature of the graphene that extends beyond the contacts, i.e. within the neighboring device, should occur. The experiment indeed shows a small increase, but the fit in this region is not as good as in the rest of the device. A possible reason for a lower temperature in the neighboring graphene could be that thermal transport in the graphene underneath the electrodes is reduced due to increased phonon scattering.

To test if thermal radiation into the open space is relevant for graphene FETs at elevated temperatures, we estimate the radiation heat loss of the graphene sheet at a uniform temperature of 800K (the highest temperature reached in the center during the actual spatially-resolved measurement) using the Stefan-Bolzmann law $I = \varepsilon \sigma A T^4$, where $\varepsilon = 2.3\%$ is the emissivity of graphene taken to be equal to the graphene absorption [30], $\sigma = 5.67 \cdot 10^{-8}$ Wm$^{-2}$K$^{-4}$ is the Bolzmann constant, and $A = (2.65 \times 1.45)$ μm$^2$ is the area of the graphene sheet. The calculated radiation heat loss is only 8nW or a negligible fraction ($10^{-6}$) of the total electric power (6.2mW) that is dissipated. Thermal radiation toward the underlying silicon-oxide can be a bit larger, because of near-field enhancement. This contribution is included, together with the thermal coupling, and the heat conduction through intervening air, in the thermal interface resistance $r_{Gr-SiO2}$.

We now turn our attention to the Raman G-phonon band. This band has also been shown to display linear temperature dependence, albeit with a smaller proportionality factor than the 2D-phonon band. [17, 31] Figure 3a shows spatial images of the G-band energy at different drain voltages. Strong phonon softening of the G-phonon band upon biasing is apparent. A striking difference from the 2D-mode case, however, is that even at zero drain bias, the G-phonon band within the graphene FET is stiffened (X$_{0\_G}$~1595cm$^{-1}$) compared to the parts of the graphene that extend beyond the device (X$_{0\_G}$~1591cm$^{-1}$). The graphene G-phonon band has been shown before to stiffen upon electrostatic gating. [18, 19] Here, unintentional doping due to the introduced high electric power densities is responsible for the persisting G-band shifts (the device has experienced several cycles of drain voltage sweeps up to -4V before these measurements). By monitoring the *I-V$_G$* characteristics of the FET, we know that this graphene sheet became more p-type over time. Presumably, the filling of electron traps in the gate oxide is responsible for the observed effect. From our Raman measurements and reference [18] we can estimate the changes in hole concentrations and Fermi level position in the graphene layer. Hole concentrations are around $p = 7 \cdot 10^{12}$ cm$^{-2}$ and the Fermi level is $E_F = -0.34$ eV in the 0V drain voltage case in Figure 3a. This compares to $p = 5 \cdot 10^{12}$ cm$^{-2}$ and $E_F = -0.29$ eV in the neighboring area of graphene that has never experienced any current flow. After the experiments in Fig. 3a were finished, the doping level within the FET had increased further to $p = 10 \cdot 10^{12}$ cm$^{-2}$ or $E_F = -0.41$ eV (Fig. 3b and 3c). The Raman-derived doping levels are consistent with doping levels estimated from the current-voltage characteristics and from the 2D/G intensity ratio (see supplemental materials). In addition they give local information, showing for example that doping levels are largest in the middle of the device where temperatures during previous device operation were highest.

Having established that charge trapping is responsible for the spatial and temporal doping variations that are revealed by the Raman G-phonon band, we must conclude that the G-band position cannot be used directly as an indication of temperature variation: First, during electric operation, additional electron or hole traps could be filled, which empty again after the drain bias is removed; And second, a drain bias might be expected to add or remove charges to the sheet more efficiently than a gate voltage on the far-away backgate does (this happens for example in short-channel MOSFETs). These effects mask the temperature effect predicted for the G-phonon band.

A valid question that can be asked is whether the 2D-band thermometry itself can be trusted. One indication, that this is indeed the case, is the very low level of doping-related shifts in the 0V-image of the 2D-phonon band in figure 1c. Since we know from the concurrently acquired G-band image (Fig. 3a) that trapped charges are present, the 2D-phonon band must be insensitive to them. Recently, A. Das et al. [18] have measured the electrolyte gate dependence of the G-phonon and the 2D-phonon bands, and they found that the energies of both Raman bands are gate dependent. However, the gate sensitivity of the 2D-phonon band is about a factor of two weaker than the G band. Furthermore, Calizo et al. [17] have found that the phonon softening due to thermal heating is about twice as strong for the 2D-phonon band. Therefore, to a good approximation, we can use the 2D-phonon band for the temperature measurement and the zero-bias G-band measurement for the dopant concentration.

To measure the carrier concentration in the active FET under drain bias, it is necessary to remove the temperature dependence of the G-phonon band, as shown in the inset of Figure 3c, by using the extracted temperatures from the concurrently-acquired 2D-phonon band. Here we assume that the measured temperatures via the 2D-band are equilibrium temperatures. Figure 3c shows the hole concentration, which decreases from $p = 10 \cdot 10^{12}$ cm$^{-2}$ at 0V drain bias to $p = 5 \cdot 10^{12}$ cm$^{-2}$ at -5 V. This corresponds to the quasi Fermi level decreasing from $E_F = -0.4$ eV to $E_F = -0.3$ eV. The reduction in hole density during application of a drain bias could indicate the presence of drain-induced barrier lowering. [32] This effect is usually observed in short-channel field-effect transistors, where the gate cannot sustain uniform doping levels in the presence of large drain voltages and the drain and source fields penetrate deep into the channel. Another possibility is that the intrinsic carrier density $n_i$ reaches or surpasses the acceptor concentration $N_A$ at the high temperatures generated by the current, which lowers the Fermi-level because charge-neutrality needs to be maintained. [33] This effect is quite common in semiconductors at elevated temperatures, but should make a smaller contribution in heavily doped graphene.

Since the carrier density in the active FET is known from our Raman measurement, we can extract the saturation velocity from the *I-V* characteristics using $j_{sat} = pev_{sat}$, where $j_{sat} = 1.614\text{mA}/1.45\mu m$ is the saturated current density and $p = 5 \cdot 10^{12} \text{cm}^{-2}$ is the hole concentration under a drain bias of -5V: $v_{sat} = 1.39 \cdot 10^7$ cm/s. Furthermore, assuming that a single inelastic scattering process due to phonons with energy $\hbar\omega$ dominates, the saturated current density is given by

$$j_{sat} = \frac{4e}{h} \hbar\omega \frac{E_F}{\pi\hbar v_F},$$

where the Fermi velocity of $v_F = 9.2 \cdot 10^7$ cm/s gives the correct number of one dimensional subbands [34] $M = \frac{E_F}{\pi\hbar v_F} W$ contributing to the current in the saturated regime, and $E_F = -0.3\text{eV}$ is the measured Fermi energy. The resulting phonon energy, $\hbar\omega = 46\text{meV}$, is well below energies of optical phonons in graphene, which are a factor of 4 larger. On the other hand, substrate polar surface phonons in silicon oxide have energies of 50meV, which is almost exactly what we measure and therefore most likely the dominant (remote) scattering process in supported graphene. This mechanism has been suggested in recent electronic transport experiments on graphene transistors [35, 36] and also in recent theoretical modeling. [37, 38]

The direct excitation of surface phonons in the silicon oxide by field coupling to the hot conduction electrons in the neighboring graphene is a process that is not included in our classical heat-flow simulation above. There we assumed that all electric energy is dissipated in the graphene and the graphene phonons couple to the $SiO_2$ phonons. Since the binding of the graphene to $SiO_2$ involves weak van-der-Waals forces, the surface of the $SiO_2$ is rough on the nanometer scale, and the optical phonon energies in graphene and $SiO_2$ don't match, it is reasonable to assume that the interface thermal resistance should be much larger than in ideal nanoscopic contacts where there is direct chemical bonding between the two materials. In contrast, the graphene/$SiO_2$ interface thermal resistance, $r_{Gr-SiO2} \sim 4 \cdot 10^{-8}$ K m$^2$/W, is within a factor of 4 of ideal nanoscopic boundaries ($r \sim 10^{-8}$ K m$^2$/W), while macroscopic boundaries, usually dealt with in processor cooling, are all much worse ($r \sim 10^{-6}$ K m$^2$/W). We suggest that the low value of the graphene-$SiO_2$ thermal resistance observed here is likely due to the remote-scattering process of hot graphene electrons. [35-38]

In conclusion, the gate stack (300nm $SiO_2$ on silicon) directly below the active graphene channel is responsible for 77% of the heat dissipation, while the remainder is carried to the graphene that extends beyond the device and metallic contacts. Operating temperatures could therefore be reduced substantially through scaling of the device (especially the gate oxide thickness), which is desirable anyway, since it allows higher integration densities and switching speeds. A surprisingly efficient thermal coupling between graphene and the underlying silicon oxide exists. We attribute part of the energy flow into the substrate to remote-scattering of 50meV substrate polar surface phonons by hot graphene electrons.

**Figure 1**

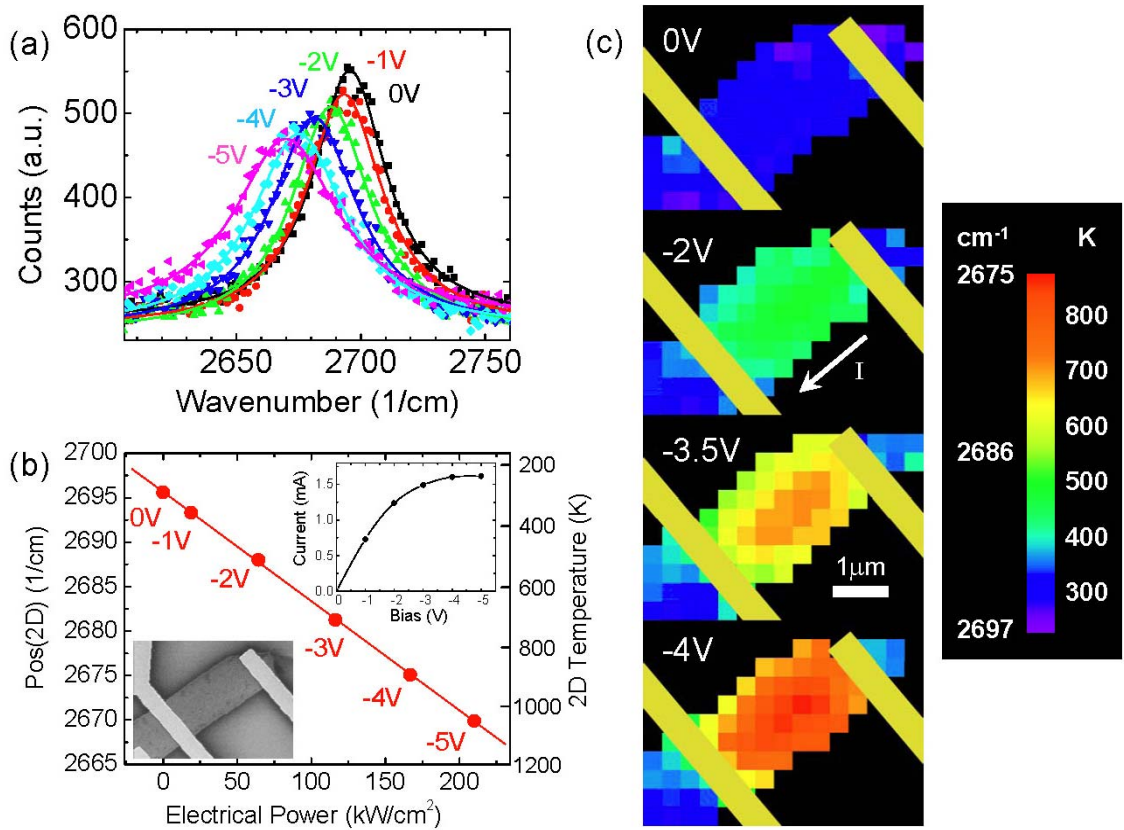

**Figure 1:** Raman 2D-phonon band of graphene under current. **(a)** 2D-band spectrum measured in the middle of the device at various drain voltages. The gate voltage was 0V, $\lambda = 514.5$ nm. Solid lines are Lorentzian fits. **(b)** Position of the 2D-phonon band as a function of applied electrical power. The solid line is a linear fit with $X_{0\_2D}$ = 2695.7 cm$^{-1}$ – 0.1229 cm$^{-1}$ · power / (kW/cm$^2$). Temperatures are calculated using the factor -29.4K/cm$^{-1}$ and defining the 0V result as room temperature. [17] **Inset (top):** Corresponding *I-V* curve. **Inset (bottom):** SEM image of the device. **(c)** Spatially-resolved images of the 2D-band position at four different drain voltages. The energies were extracted at each pixel by a Lorentzian fit. The graphene flake extends beyond the left and right contacts (indicated in yellow). The 2D-band images were taken concurrently with the G-band images in Figure 3a below.

**Figure 2**

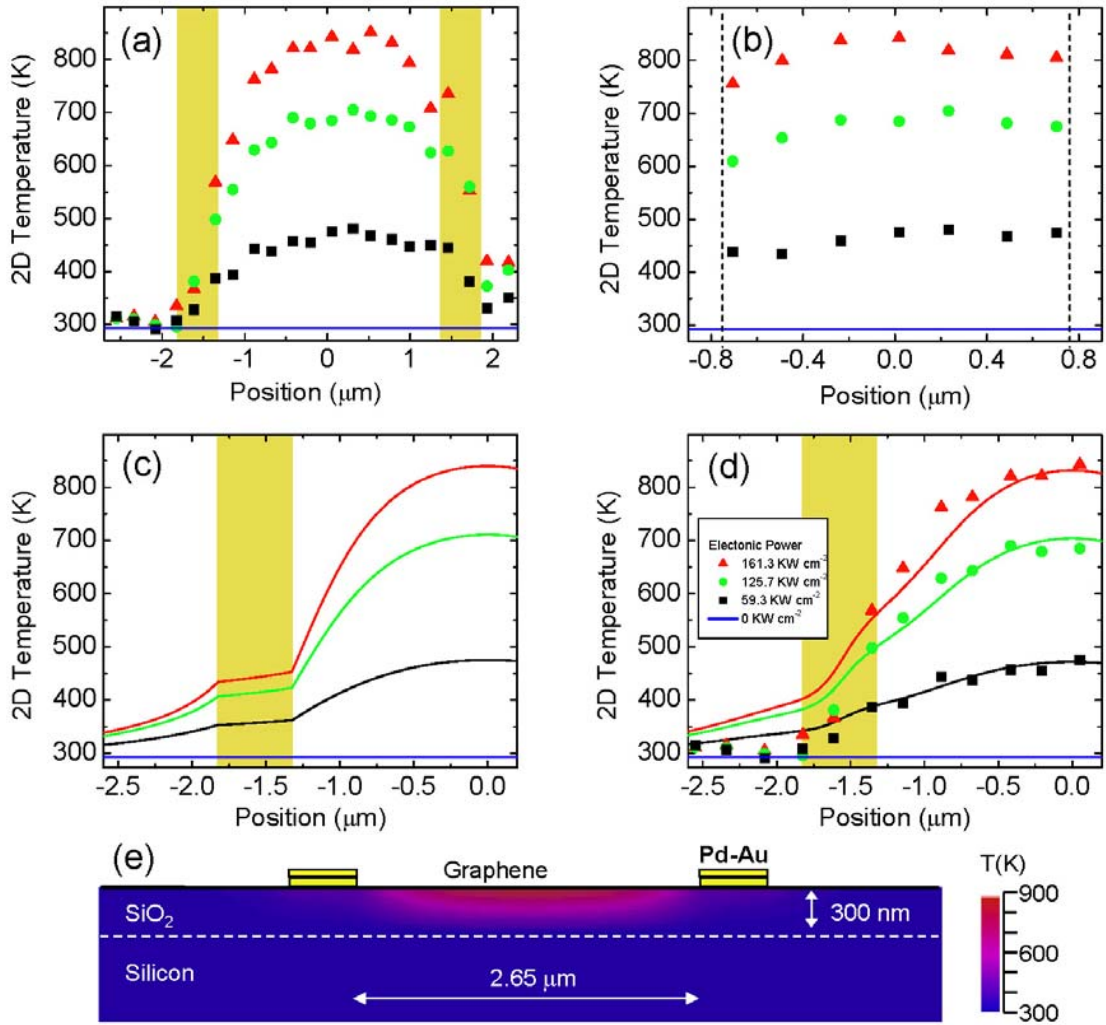

**Figure 2:** Model of the temperature distribution. **(a)** Experimental 2D-band temperature along the graphene sheet (parallel to the current flow) for 59.3 KW cm$^{-2}$ (black), 125.7 KW cm$^{-2}$ (green), and 161.3 KW cm$^{-2}$ (red) dissipated electronic power. The blue line corresponds to the zero power case, defined as room temperature. Electrodes are depicted in yellow. **(b)** Experimental 2D-band temperature profile across the graphene sheet (perpendicular to the current flow) for the same biasing conditions. The dotted lines mark the ends of the graphene sheet. **(c)** Modeled temperature profile along the graphene device for the same biasing conditions. **(d)** Comparison of the model with the experimental 2D-band temperature. The best overall fit was achieved for a thermal interface resistance of $r_{Gr-SiO2} = 4.2 \cdot 10^{-8}$ K m²/W between graphene and the substrate. The modeled curves from (c) were broadened by the laser spot size of 0.5 μm. **(e)** Cross section of the temperature distribution in the gate stack for 161.3 KW cm$^{-2}$ of dissipated electronic power.

**Figure 3**

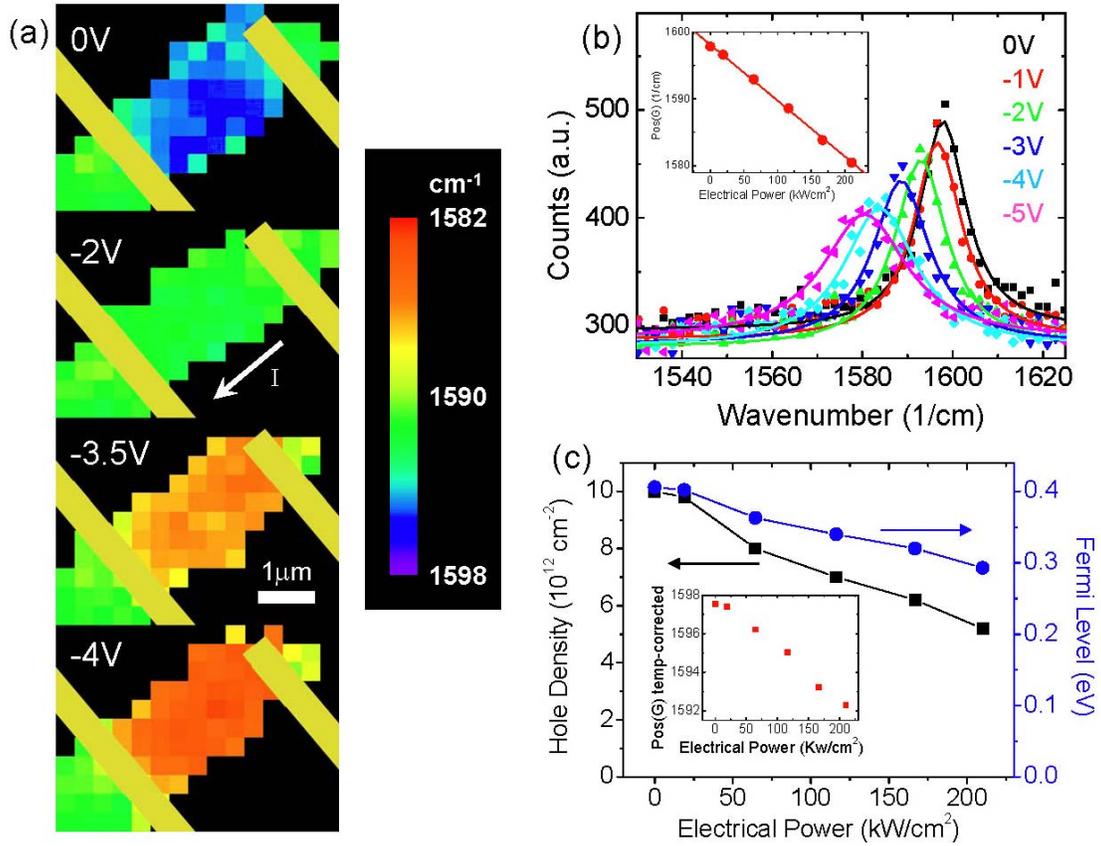

**Figure 3:** Raman G-phonon band of graphene under current. **(a)** Spatially-resolved images of the G-band position at four different drain voltages. The gate voltage was 0V. The G-phonon band is very sensitive to both, the increase in temperature, as well as trapped charges in the oxide. All these images were taken after initial electrical measurements up to -4V drain voltage, which led to the trapped charges visible in the 0V image of Fig. 3a. **(b)** Spectra of the Raman G-phonon band in the middle of the device under drain bias. (Gate voltage 0V, $\lambda$ = 514.5 nm, solid lines Lorentzian fits). This measurement was done after the spatially resolved Fig. 3a. **Inset:** G-band position as a function of applied electronic power. The solid line is a linear fit with $X_{0\_G}$ = 1598.1 cm$^{-1}$ – 0.0836 cm$^{-1}$ · power / (kW/cm$^2$). **(c)** Hole density (black squares) and Fermi level (blue circles) as a function of applied electronic power. **Inset:** Temperature-corrected G-band

position as a function of dissipated electronic power. The temperature component in the G-band position has been subtracted, using the 2D-derived temperature.

# Energy dissipation in graphene field-effect transistors

Marcus Freitag*, Mathias Steiner, Yves Martin, Vasili Perebeinos,
Zhihong Chen, James C. Tsang, Phaedon Avouris

IBM TJ Watson Research Center, Yorktown Heights, NY 10591
* email: mfreitag@us.ibm.com

**Supplement**

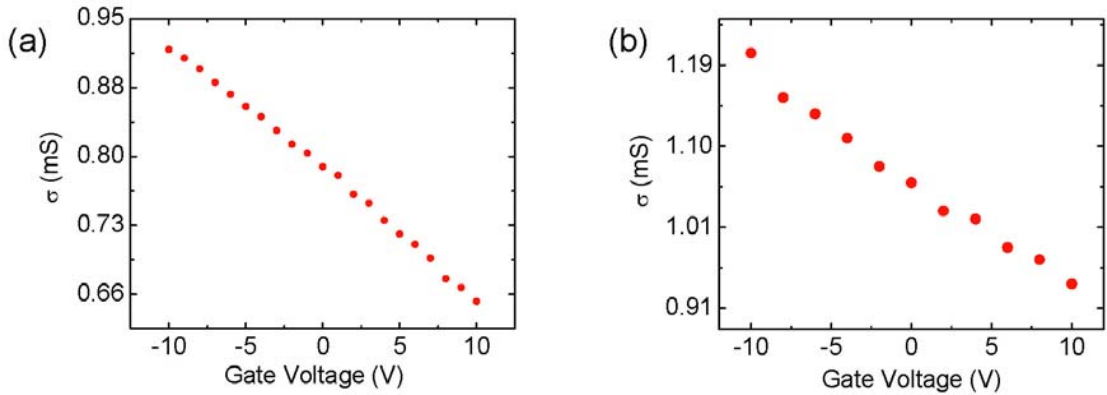

**Supplement 1:** Calculation of the charge density using the gate-voltage characteristics. **(a)** Initial Gate-voltage characteristics of the graphene device. The conductivity $\sigma = \frac{L}{W} \cdot G$ is plotted. (Length L=2.65μm and width W=1.45μm). The data was taken at $V_D$=-10mV. A linear fit to the data yields $\sigma = 0.7899 \text{ mS} - 0.0135 \frac{\text{mS}}{\text{V}} \cdot V_G$, suggesting that the Dirac point is in the vicinity of $V_G = +59$ V. Using $p = \frac{C}{e} V_G$ and the estimated backgate capacitance $C = 1.2 \cdot 10^{-8}$ Fcm$^{-2}$, we get a hole density of $p = 4.4 \cdot 10^{12}$ cm$^{-2}$, consistent with the Raman result of $p = 5 \cdot 10^{12}$ cm$^{-2}$ obtained for the neighboring graphene in Fig. 3a that has never seen any current flow.

**(b)** Gate-voltage characteristics of the graphene device taken after initial electrical measurements up to $V_D$=-4V, but prior to the Raman measurements in the manuscript. The data was taken at $V_D$=-400mV. A linear fit to the data yields $\sigma = 1.0592$ mS $- 0.0124 \frac{\text{mS}}{\text{V}} \cdot V_G$, suggesting that the Dirac point has now moved to $V_G = +85$ V. Again, using $p = \frac{C}{e} V_G$ and the estimated backgate capacitance $C = 1.2 \cdot 10^{-8}$ Fcm$^{-2}$, we get a hole density of $p = 6.4 \cdot 10^{12}$ cm$^{-2}$, consistent with the Raman result of $p = 7 \cdot 10^{12}$ cm$^{-2}$ taken in the middle of the active graphene sheet in Fig. 3a, after initial electrical measurements up to $V_D$=-4V.

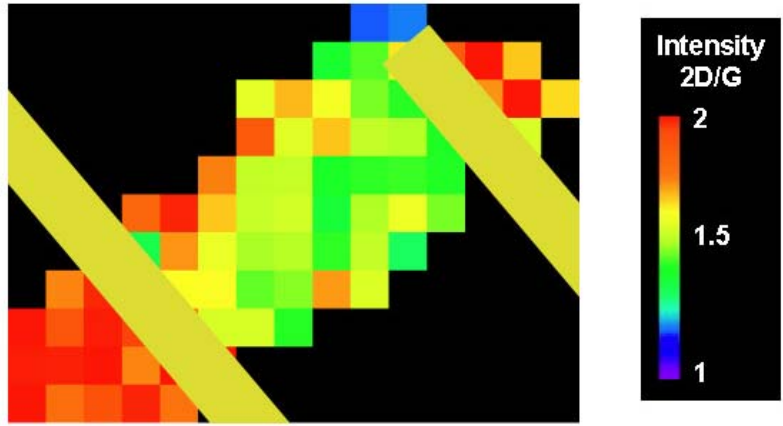

**Supplement 2:** Spatially-resolved intensity ratio of the 2D and G bands for 0V drain bias. In the center of the device, the ratio is around 1.5, whereas in the neighboring graphene the ratio is close to 2. According to reference [18] this would correspond to a doping level of $p = 7 \cdot 10^{12}$ cm$^{-2}$ within the device and $p = 4.5 \cdot 10^{12}$ cm$^{-2}$ in the neighboring graphene, consistent with the Raman results from the G-band shift of $p = 7 \cdot 10^{12}$ cm$^{-2}$ and $p = 5 \cdot 10^{12}$ cm$^{-2}$ respectively (Fig. 3a).

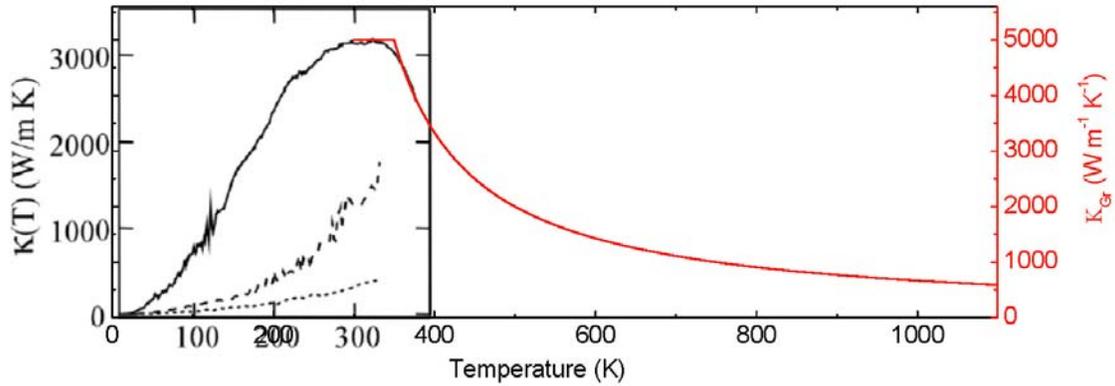

**Supplement 3:** Assumed temperature dependence of the thermal conductivity of graphene at elevated temperatures. The solid black curve between T=10K and 380K is taken from a measurement on multi-walled carbon nanotubes. [29] The red curve is used to model the temperature dependence of the graphene thermal conductivity. Note that the room-temperature thermal conductivity of the nanotube, $\kappa_{CNT} \sim 3000$ Wm$^{-1}$K$^{-1}$, is not the same as the reported graphene thermal conductivity at room temperature, $\kappa_{Gr} = 5000$ Wm$^{-1}$K$^{-1}$, [24] and we re-scale the MWNT curve to coincide with the graphene result at *T*=300K.

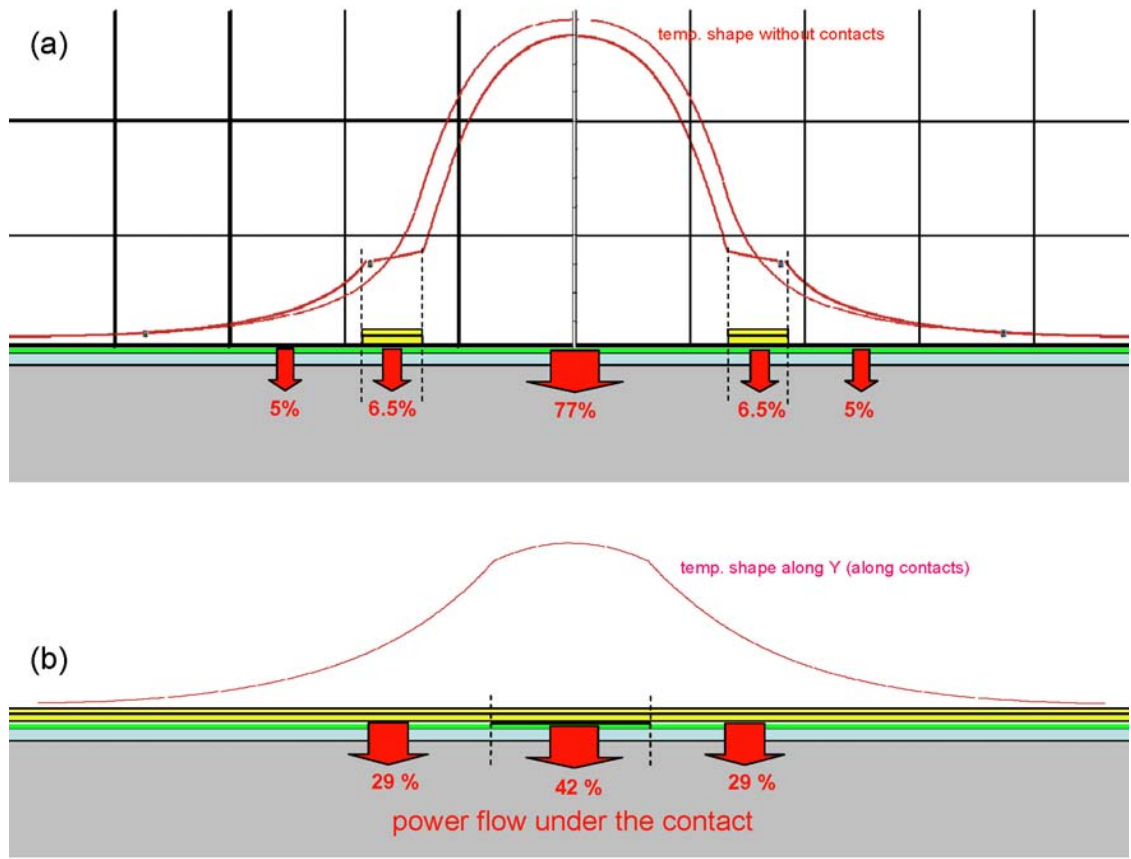

**Supplement 4:** Heat dissipation pathways. **(a)** Temperature profiles, modeled with and without thermal transport in the metal contacts for an electrical power density of 161.3 KW cm$^{-2}$. Through each of the contacts 6.5% of the total electrical power is dissipated. Most of the power (77%) is dissipated through the SiO$_2$ below the graphene. **(b)** Modeled temperature profile in the metallic leads in the y direction (perpendicular to the device).